\documentstyle[11pt]{article}
\normalbaselines
\begin{document}
\bibliographystyle{unsrt}

\vbox{\vspace{38mm}}
\begin{center}
{\LARGE \bf Lie Symmetries of Yang-Mills Equations}\\[5mm]

Louis Marchildon\\{\it D\'{e}partement de physique,
Universit\'{e} du
Qu\'{e}bec,\\Trois-Rivi\`{e}res, Qu\'{e}bec, Canada G9A
5H7\\(e-mail:marchild@uqtr.uquebec.ca)}\\[5mm]
(Submitted 13 July 1995)\\[5mm] \end{center}

\begin{abstract} We investigate Lie symmetries of general
Yang-Mills equations.
For this purpose, we first write down the second prolongation of
the symmetry
generating vector fields, and compute its action on the
Yang-Mills equations.
Determining equations are then obtained, and solved completely.
Provided that
Yang-Mills equations are locally solvable, this allows for a
complete
classification of their Lie symmetries.  Symmetries of Yang-Mills
equations in
the Lorentz gauge are also investigated.  PACS: 02.20.+b
\end{abstract}

\section{Introduction}
Consider a system of {\it n}-th order nonlinear partial
differential equations
for a number of independent variables $x$ and dependent variables
$A$:
\begin{equation}
\Delta _{\nu} (x,A,\partial A,\ldots ,\partial ^{(n)} A) = 0.
\label{delta} \end{equation}
By a symmetry of this system, we shall mean any mapping of the
independent and
dependent variables that transforms an arbitrary solution of
(\ref{delta}) into
a solution.  A Lie symmetry is a symmetry that belongs to a local
Lie group of
transformations.  A Lie symmetry is generated by a differential
operator $v$
which is a linear combination of partial derivatives with respect
to the $x$
and $A$.  In general, the coefficients of the linear combination
depend on $x$
and $A$.

There is a well-defined method for the determination of all Lie
symmetries of
Eq.~(\ref{delta})\@.
It involves the computation of the so-called {\it n}-th
prolongation of $v$, denoted by $\mbox{pr} ^{(n)} v$.  The {\it
n}-th
prolongation of $v$ is a linear combination of partial
derivatives with respect
to $x$, $A$, and with respect to all partial derivatives of $A$
up to the {\it
n}-th order.  One can show that, provided Eq.~(\ref{delta})
is locally
solvable and has maximal rank, $v$ generates a symmetry of
(\ref{delta}) if and
only if the following holds~\cite{Olver}:
\begin{equation}
\left [\mbox{pr} ^{(n)} v \right ] \Delta _{\nu} = 0 \mbox{
whenever } \Delta
_{\nu} = 0.
\label{prdelta} \end{equation}
This constitutes a system of linear equations for the
coefficients of partial
derivatives making up the operator $v$.

In this paper, we shall investigate Lie symmetries of general
Yang-Mills
equations.  Such equations are characterized by local gauge
invariance under a
compact semisimple Lie group.  Their importance can hardly be
overestimated, since they form the basis of current theories of
the strong and electroweak interactions.  Yet, as pointed out in
Ref.~\cite{Torrea}, the techniques of symmetry analysis have not
been applied systematically to the Yang-Mills equations.
Previous work centered on the SU(2) case, for which Lie
symmetries of Yang-Mills
equations in the Lorentz gauge, and of so-called self-dual
Yang-Mills equations, have been obtained~\cite{Rosen,Ker}.

Just recently, however, Torre investigated what he calls natural
symmetries of general Yang-Mills equations~\cite{Torre}.
They are
generalized symmetries (in the sense of Ref.~\cite{Olver}) that,
roughly speaking, have a simple behavior under Poincar\'{e} and
gauge transformations of the fields.  Torre showed that all such
symmetries come from the gauge transformations admitted by the
equations.

In this paper we shall show that, provided the Yang-Mills
equations are locally solvable, their Lie symmetries all come
from local gauge transformations and conformal transformations.
Our results are thus consistent with Torre's.  In one aspect they
are less general, since we investigate Lie instead of generalized
symmetries.  In another aspect, however, they are more general,
since we do not assume any specific behavior under Poincar\'{e}
and gauge transformations.  We should also point out that Torre
makes use of the spinor formalism, applicable to four-dimensional
manifolds.  Our method can straightforwardly be adapted to
higher-dimensional manifolds.

In Section~2, we write down general Yang-Mills equations, the
prolongation
formulas, and discuss the question of local solvability.
Determining equations
associated with Yang-Mills equations are obtained in Section~3,
and completely solved in Section~4.
Yang-Mills equations in the Lorentz gauge are investigated in
Section~5.

\section{Yang-Mills Equations}
With suitable choice of units, Yang-Mills equations can be
written
as~\cite{Abers,Huang}
\begin{equation}
\partial _{\mu} F_a^{\mu\nu} + C_{abc} A_{b \mu} F_c^{\mu\nu} =
0,
\label{YM} \end{equation}
where
\begin{equation}
F_a^{\mu\nu} =  \partial ^{\mu} A_a^{\nu} - \partial ^{\nu}
A_a^{\mu} + C_{abc}
A_b^{\mu} A_c^{\nu}.
\label{F} \end{equation}
Here $\partial _{\mu}$ represents a partial derivative with
respect to the
independent variable $x^{\mu} (\mu = 0, 1, 2, 3)$\@.  The
$A_b^{\mu}$ are
dependent variables.  Greek indices, associated with space-time,
are raised and
lowered with the Minkowski metric $g_{\mu\nu}$, with signature
$(+1, -1, -1, -
1)$\@.  Latin indices are associated with the structure constants
$C_{abc}$ of
a compact semisimple Lie group.  The $C_{abc}$ can always be
chosen so that
they are completely antisymmetric and satisfy
\begin{equation}
C_{acd} C_{bcd} = \delta _{ab},
\label{Kro} \end{equation}
where $\delta _{ab}$ is the Kronecker delta.

Eliminating $F$ from Eqs.\ (\ref{YM}) and (\ref{F}), we find that
the Yang-Mills equations can be written as
\begin{eqnarray}
\lefteqn{\partial _{\mu} \partial ^{\mu} A_a^{\nu} - \partial
_{\mu} \partial ^{\nu} A_a^{\mu} + 2 C_{abc} A_b^{\mu} \partial
_{\mu} A_c^{\nu} + C_{abc} (\partial _{\mu} A_b^{\mu}) A_c^{\nu}}
\nonumber \\
& & \mbox{} - C_{abc} A_{b\mu} \partial ^{\nu} A_c^{\mu} +
C_{abc} C_{cdl} A_d^{\mu} A_l^{\nu} A_{b\mu} = 0.
\label{YMX} \end{eqnarray}
The Yang-Mills equations are second-order nonlinear partial
differential
equations.  Generators of symmetry transformations are given by
\begin{equation}
v = H^\kappa \partial _{\kappa} + \Phi _d ^{\kappa} \frac
{\partial} {\partial
A_d^{\kappa}},
\label{v} \end{equation}
where $H^\kappa$ and $\Phi _d ^{\kappa}$ are functions of
$x^{\mu}$ and
$A_a^{\nu}$\@.  The second prolongation of $v$ is given by
\begin{equation}
\mbox{pr} ^{(2)} v = H^\kappa \partial _{\kappa} + \Phi _d
^{\kappa} \frac
{\partial} {\partial A_d^{\kappa}} + \Phi _{d\lambda} ^{\kappa}
\frac
{\partial} {\partial (\partial _{\lambda} A_d^{\kappa})} + \Phi
_{d(\lambda
\pi)} ^{\kappa} \frac {\partial} {\partial (\partial _{\lambda}
\partial _{\pi}
A_d^{\kappa})}.
\label{pr2} \end{equation}
The parentheses around $\lambda$ and $\pi$ indicate that the
implicit summation
is restricted to values of the indices such that $\lambda \leq
\pi$, that is,
over distinct partial derivatives only.  The coefficients $\Phi
_{d\lambda}
^{\kappa}$ and $\Phi _{d\lambda \pi} ^{\kappa}$ are functions of
$x^{\mu}$ and
$A_a^{\nu}$\@.  They are given by~\cite {Olver}
\begin{equation}
\Phi _{d\lambda} ^{\kappa} = D_{\lambda} \left\{ \Phi _d^{\kappa}
- H^{\mu}
\partial _{\mu} A_d^{\kappa} \right\} + H^{\mu} \partial
_{\lambda} \partial
_{\mu} A_d^{\kappa},
\label{Phi3} \end{equation}
\begin{equation}
\Phi _{d\lambda \pi} ^{\kappa} = D_{\pi} D_{\lambda} \left\{ \Phi
_d^{\kappa} -
H^{\mu} \partial _{\mu} A_d^{\kappa} \right\} + H^{\mu} \partial
_{\pi}
\partial _{\lambda} \partial _{\mu} A_d^{\kappa}.
\label{Phi4} \end{equation}
The operator $D_{\lambda}$ and $D_{\pi}$ are total derivatives,
that is,
\begin{equation}
D_{\lambda} = \partial_{\lambda} + (\partial _{\lambda}
A_n^{\alpha})
\frac{\partial}{ \partial A_n^{\alpha}} + (\partial _{\lambda}
\partial
_{\beta} A_n^{\alpha}) \frac{\partial}{ \partial (\partial
_{\beta}
A_n^{\alpha})} + (\partial _{\lambda} \partial _{(\beta} \partial
_{\gamma )}
A_n^{\alpha}) \frac{\partial}{ \partial (\partial _{\beta}
\partial _{\gamma}
A_n^{\alpha})}.
\label{TD} \end{equation}
Note that, in the last term, summation is restricted to values of
the
indices such that $\beta  \leq \gamma$.

In this paper, we shall let Eq.~(\ref{delta}) represent the
Yang-Mills equations, and investigate the most general functions
$H^\kappa$ and
$\Phi _d ^{\kappa}$ that satisfy Eq.~(\ref{prdelta}).
As shown in Ref.~\cite{Olver},
Eq.~(\ref{prdelta})
is a sufficient condition for $v$ to generate a symmetry.
Provided the
Yang-Mills equations are locally solvable and have maximal rank,
this is also a
necessary condition.

It is easy to check that the Yang-Mills equations have maximal
rank in the
sense of Ref.~\cite{Olver}.
It is less easy to prove that they are locally
solvable.
Local solvability means that one can find solutions for arbitrary
values of the
partial derivatives at a given point, compatible with the
equations.  A
sufficient (but not necessary) condition for local solvability is
that the
equations be in (general) Kovalevskaya form.  Unfortunately, the
Yang-Mills
equations are not in that form.  Nevertheless, it is likely that
the Yang-Mills
equations are locally solvable.  As pointed out in
Ref.~\cite{Olver}, the
main reason why
an analytic system is not locally solvable is the existence of
additional
constraints on partial derivatives implied by the equations
themselves.  It
might appear that such additional constraints could be put on the
Yang-Mills
fields by acting on Eq.~(\ref{YMX})
with the operator $\partial _{\nu}$.  The
first two terms then vanish, yielding an equation for the
functions $A_b
^{\mu}$ and their partial derivatives.  But we show, in Appendix
A, that the
resulting equation holds as an identity.  Therefore, that
operation gives no
additional constraints on the fields.

A word on notations.  It has already been said that parentheses
enclosing a
pair of indices indicate that the implicit summation should be
carried out only
on distinct pairs of indices.  It has also been assumed that
the implicit summation convention, on Greek as well as Latin
indices, is
effective.  There will, however, be instances where we will not
want to sum
over repeated indices.  Obviously, we could just put summation
signs where
needed, and no such signs elsewhere.  However, the summation
convention is so
useful that it is better to proceed otherwise.  We shall use the
summation
convention on repeated indices, {\it unless} indices have a
caret, in which
case no summation will be carried out. This means, for instance,
that in an
equation like
\begin{equation}
M^{\mu}_{\mu} = N^{\hat{\alpha}}_{\hat{\alpha}} ,
\end{equation}
summation is carried out over $\mu$ but not over $\hat{\alpha}$,
the latter
index having a specific value.

For later purposes, it is useful to write down the
Yang-Mills equations (\ref{YMX}) in a form that exhibits each
second-order partial derivative.  For each value of the index
$a$, there are
four equations, corresponding to each value of the index $\nu$\@.
They are given by
\begin{eqnarray}
\lefteqn{\partial _1 \partial _1 A_a^0 = - \partial _2 \partial
_2 A_a^0 - \partial _3 \partial _3 A_a^0 - \partial _1 \partial
_0 A_a^1 - \partial _2 \partial _0
A_a^2 - \partial _3 \partial _0 A_a^3} \nonumber \\
 & & \mbox{} + 2 C_{abc} A_b^{\mu}
\partial _{\mu} A_c^0 + C_{abc} (\partial _{\mu} A_b^{\mu}) A_c^0
- C_{abc} A_{b\mu} \partial ^0 A_c^{\mu} + C_{abc} C_{cdl}
A_d^{\mu} A_l^0 A_{b\mu}.
\label{YM0} \end{eqnarray}
\begin{eqnarray}
\lefteqn{\partial _2 \partial _2 A_a^1 = - \partial _3 \partial
_3 A_a^1 + \partial _0 \partial _0 A_a^1 + \partial _2 \partial
_1 A_a^2 + \partial _3 \partial _1 A_a^3 + \partial _0 \partial
_1 A_a^0} \nonumber \\
 & & \mbox{} + 2 C_{abc} A_b^{\mu}
\partial _{\mu} A_c^1 + C_{abc} (\partial _{\mu} A_b^{\mu}) A_c^1
- C_{abc} A_{b\mu} \partial ^1 A_c^{\mu} + C_{abc} C_{cdl}
A_d^{\mu} A_l^1 A_{b\mu}.
\label{YM1} \end{eqnarray}
\begin{eqnarray}
\lefteqn{\partial _1 \partial _1 A_a^2 = - \partial _3 \partial
_3 A_a^2 + \partial _0 \partial _0 A_a^2 + \partial _1 \partial
_2 A_a^1 + \partial _3 \partial _2 A_a^3 + \partial _0 \partial
_2 A_a^0} \nonumber \\
 & & \mbox{} + 2 C_{abc} A_b^{\mu}
\partial _{\mu} A_c^2 + C_{abc} (\partial _{\mu} A_b^{\mu}) A_c^2
- C_{abc} A_{b\mu} \partial ^2 A_c^{\mu} + C_{abc} C_{cdl}
A_d^{\mu} A_l^2 A_{b\mu}.
\label{YM2} \end{eqnarray}
\begin{eqnarray}
\lefteqn{\partial _1 \partial _1 A_a^3 = - \partial _2 \partial
_2 A_a^3 + \partial _0 \partial _0 A_a^3 + \partial _1 \partial
_3 A_a^1 + \partial _2 \partial _3
A_a^2 + \partial _0 \partial _3 A_a^0} \nonumber \\
& & \mbox{} + 2 C_{abc} A_b^{\mu}
\partial _{\mu} A_c^3 + C_{abc} (\partial _{\mu} A_b^{\mu}) A_c^3
- C_{abc} A_{b\mu} \partial ^3 A_c^{\mu} + C_{abc} C_{cdl}
A_d^{\mu} A_l^3 A_{b\mu}.
\label{YM3} \end{eqnarray}

\section{Determining Equations}
In this section, we will translate condition (\ref{prdelta}) for
the Yang-Mills case in explicit equations.  First, we have to
compute the
coefficients $\Phi _{d\lambda} ^{\kappa}$ and $\Phi _{d\lambda
\pi} ^{\kappa}$
that appear in the prolongation formula (\ref{pr2})\@.
Substituting
Eq.~(\ref{TD}) into Eqs.\ (\ref{Phi3}) and (\ref{Phi4}), we find
\begin{equation}
\Phi _{d\lambda}^{\kappa} = \partial _{\lambda} \Phi _d^{\kappa}
- (\partial
_{\lambda} H^{\mu}) \partial _{\mu} A_d^{\kappa} + (\partial
_{\lambda}
A_n^{\alpha}) \frac{\partial \Phi _d^{\kappa}}{\partial
A_n^{\alpha}} -
(\partial _{\lambda} A_n^{\alpha}) (\partial _{\mu} A_d^{\kappa}
)
\frac{\partial H^{\mu}}{\partial A_n^{\alpha}},
\label{Phi3X}\end{equation}
\begin{eqnarray}
\lefteqn{\Phi _{d \lambda \pi}^{\kappa} = \partial _{\pi}
\partial _{\lambda} \Phi
_d^{\kappa} - (\partial _{\mu} A_d^{\kappa}) \partial _{\pi}
\partial _{\lambda} H^{\mu} + (\partial _{\lambda} A_n^{\alpha})
\frac{ \partial}{
\partial A_n^{\alpha}} \partial _{\pi} \Phi _d^{\kappa} +
(\partial _{\pi}
A_n^{\alpha}) \frac{ \partial}{ \partial A_n^{\alpha}} \partial
_{\lambda} \Phi _d^{\kappa}} \nonumber \\
& & \mbox{} - (\partial _{\lambda} A_n^{\alpha}) (\partial _{\mu}
A_d^{\kappa}) \frac{
\partial}{ \partial A_n^{\alpha}} \partial _{\pi} H^{\mu} -
(\partial _{\pi}
A_n^{\alpha}) (\partial _{\mu} A_d^{\kappa}) \frac{ \partial}{
\partial
A_n^{\alpha}} \partial _{\lambda} H^{\mu} + (\partial _{\pi}
A_p^{\beta})
(\partial _{\lambda} A_n^{\alpha}) \frac{ \partial ^2 \Phi
_d^{\kappa}}{
\partial A_p^{\beta} \partial A_n^{\alpha}} \nonumber \\
& & \mbox{} - (\partial _{\pi} \partial _{\mu} A_d^{\kappa})
\partial _{\lambda}
H^{\mu}  - (\partial _{\lambda} \partial _{\mu} A_d^{\kappa})
\partial _{\pi}
H^{\mu} + (\partial _{\pi} \partial _{\lambda} A_n^{\alpha})
\frac{ \partial
\Phi _d^{\kappa}}{ \partial A_n^{\alpha}} \nonumber \\
& & \mbox{} - (\partial _{\pi} A_p^{\beta}) (\partial _{\lambda}
A_n^{\alpha})
(\partial _{\mu} A_d^{\kappa}) \frac{ \partial ^2 H^{\mu}}{
\partial
A_p^{\beta} \partial A_n^{\alpha}} - (\partial _{\pi}
A_n^{\alpha}) (\partial
_{\lambda} \partial _{\mu} A_d^{\kappa}) \frac{ \partial
H^{\mu}}{ \partial
A_n^{\alpha}} \nonumber \\
& & \mbox{} - (\partial _{\lambda} A_n^{\alpha}) (\partial _{\pi}
\partial _{\mu}
A_d^{\kappa}) \frac{ \partial H^{\mu}}{ \partial A_n^{\alpha}} -
(\partial
_{\mu} A_d^{\kappa}) (\partial _{\pi} \partial _{\lambda}
A_n^{\alpha}) \frac{
\partial H^{\mu}}{ \partial A_n^{\alpha}} .
\label{Phi4X} \end{eqnarray}
We note that restrictions on summations have disappeared.
Applying the
prolongation operator (\ref{pr2}) to Eq.~(\ref{YMX}), we obtain
\begin{eqnarray}
\lefteqn{\Phi _d^{\kappa} C_{adc} (2 \partial _{\kappa} A_c^{\nu}
- \partial ^{\nu}
A_{c \kappa} ) + \Phi _d^{\nu} C_{abd} \partial _{\mu} A_b^{\mu}
+ \Phi _d^{\kappa} (C_{abc} C_{cdl} + C_{adc} C_{cbl} ) A_l^{\nu}
A_{b \kappa}} \nonumber \\
& & \mbox{} + \Phi _d^{\nu} C_{abc} C_{cld} A_l^{\mu} A_{b \mu} +
\Phi ^{\nu} _{d
\lambda} 2 C_{abd} A_b^{\lambda} + \Phi ^{\kappa} _{d \kappa}
C_{adc} A_c^{\nu}
- \Phi _d ^{\kappa \nu} C_{abd} A_{b \kappa} + {{\Phi _a^{\nu}}
_{\lambda}}
^{\lambda} - {{\Phi _a^{\lambda}} _{\lambda}} ^{\nu} = 0.
\label{Sym} \end{eqnarray}
Again, restrictions on summations have disappeared.

We now substitute Eqs.\ (\ref{Phi3X}) and (\ref{Phi4X})
into (\ref{Sym}).  Regrouping coefficients of various derivatives
of $A$, we
get
\begin{eqnarray}
\lefteqn{\left\{ \partial _{\lambda} \partial ^{\lambda} \Phi
_a^{\nu} - \partial
_{\lambda} \partial ^{\nu} \Phi _a^{\lambda} + A_b^{\lambda}
\left[ 2 C_{abd} \partial _{\lambda} \Phi _d^{\nu} - g_{\lambda}
^{\nu} C_{abd}  \partial _{\kappa} \Phi _d^{\kappa} - C_{abd}
\partial ^{\nu} \Phi _{d \lambda} \right] \right.} \nonumber \\
& & \mbox{} + \left. A_l^{\mu} A_ {b \kappa} \left[ g_{\mu}
^{\nu} ( C_{abc} C_{cdl} +
C_{adc} C_{cbl} ) \Phi _d ^{\kappa} + g_{\mu} ^{\kappa} C_{abc}
C_{cld} \Phi
_d^{\nu} \right] \right\} \nonumber \\
& & \mbox{} + (\partial _{\lambda} A_n^{\alpha}) \left\{ 2
g_{\alpha}^{\nu} C_{adn}
\Phi_d ^{\lambda} + g ^{\lambda \nu} C_{and} \Phi_{d \alpha} +
g_{\alpha}^{\lambda} C_{and} \Phi_d ^{\nu} - \delta _{an} g
_{\alpha} ^{\nu}
\partial _{\mu} \partial ^{\mu} H^{\lambda} + \delta _{an}
\partial _{\alpha}
\partial ^{\nu} H^{\lambda} \right. \nonumber \\
& & \mbox{} + 2 \frac{ \partial}{ \partial A_n ^{\alpha}}
\partial ^{\lambda} \Phi _a
^{\nu} - \frac{ \partial}{ \partial A_n ^{\alpha}} \partial
^{\nu} \Phi _a
^{\lambda} - g^{\nu \lambda} \frac{ \partial}{ \partial A_n
^{\alpha}} \partial
_{\mu} \Phi _a ^{\mu} \nonumber \\
& & \mbox{} + A_b^{\mu} \left[ - 2 \delta _{nd} g_{\alpha} ^{\nu}
C_{abd} \partial
_{\mu} H^{\lambda} + g_{\mu} ^{\nu} C_{abn} \partial _{\alpha}
H^{\lambda} +
g_{\mu \alpha} C_{abn} \partial ^{\nu} H^{\lambda} \right.
\nonumber \\
& & \mbox{} + \left. \left. 2 g_{\mu} ^{\lambda} C_{abd} \frac{
\partial \Phi _d
^{\nu}}{
\partial A_n ^{\alpha}} - g_{\mu}^{\nu} C_{abd} \frac{ \partial
\Phi _d
^{\lambda} }{ \partial A_n ^{\alpha}} - g^{\nu \lambda} C_{abd}
\frac{ \partial
\Phi _{d \mu}}{ \partial A_n ^{\alpha}} \right] \right\}
\nonumber \\
& & \mbox{} + (\partial _{\kappa} A_n^{\alpha}) (\partial
_{\lambda} A_p^{\beta})
\left\{ g^{\kappa \lambda} \frac{ \partial ^2 \Phi _a^{\nu}}{
\partial A_p
^{\beta} \partial A_n ^{\alpha}} -  g^{\nu \lambda} \frac{
\partial ^2 \Phi
_a^{\kappa}}{ \partial A_p ^{\beta} \partial A_n ^{\alpha}}
\right. \nonumber
\\
& & \mbox{} + \delta _{an} g_{\alpha} ^{\lambda} \frac{
\partial}{ \partial A_p
^{\beta}} \partial ^{\nu} H^{\kappa} - 2 \delta _{an} g_{\alpha}
^{\nu} \frac{
\partial}{ \partial A_p ^{\beta}} \partial ^{\lambda} H^{\kappa}
+ \delta _{ap}
g ^{\nu \kappa} \frac{ \partial}{ \partial A_n ^{\alpha}}
\partial _{\beta}
H^{\lambda} \nonumber \\
& & \mbox{} + \left. A_b ^{\mu} \left[ - 2 g_{\mu} ^{\kappa} g
_{\beta} ^{\nu} C_{abp}
\frac{ \partial H^{\lambda}}{ \partial A_n ^{\alpha}} + g_{\mu}
^{\nu} g
_{\beta} ^{\kappa} C_{abp} \frac{ \partial H^{\lambda}}{ \partial
A_n ^{\alpha}}  + g_{\mu \beta} g ^{\nu \kappa} C_{abp} \frac{
\partial H^{\lambda}}{ \partial A_n ^{\alpha}} \right] \right\}
\nonumber \\
& & \mbox{} + (\partial _{\kappa} A_n^{\alpha}) (\partial
_{\lambda} A_p^{\beta})
(\partial _{\mu} A_a ^{\pi}) \left\{g^{\nu \lambda} g_{\pi}
^{\kappa} \frac{
\partial ^2 H^{\mu}}{ \partial A_p ^{\beta} \partial A_n
^{\alpha}} - g^{\kappa
\lambda} g_{\pi} ^{\nu} \frac{ \partial ^2 H^{\mu}}{ \partial A_p
^{\beta}
\partial A_n ^{\alpha}} \right\} \nonumber \\
& & \mbox{} + (\partial _{\lambda} \partial _{\mu} A_n ^{\alpha})
\left\{ - 2 \delta
_{an} g_{\alpha} ^{\nu} \partial ^{\lambda} H ^{\mu} + \delta
_{an} g ^{\lambda
\nu} \partial _{\alpha} H ^{\mu} + \delta _{an} g _{\alpha}
^{\lambda} \partial
^{\nu} H ^{\mu} + g^{\lambda \mu} \frac{ \partial \Phi _a
^{\nu}}{ \partial A_n
^{\alpha}}  - g^{\nu \mu} \frac{ \partial \Phi _a ^{\lambda}}{
\partial A_n
^{\alpha}} \right\} \nonumber \\
& & \mbox{} + (\partial ^{\kappa} A_n ^{\alpha})(\partial
_{\lambda} \partial _{\mu}
A_p ^{\beta}) \left\{ - 2 \delta _{ap} g_{\kappa} ^{\lambda}
g_{\beta} ^{\nu}
\frac{ \partial H^{\mu}}{ \partial A_n ^{\alpha}} + \delta _{ap}
g_{\kappa}
^{\nu} g_{\beta} ^{\lambda} \frac{ \partial H^{\mu}}{ \partial
A_n ^{\alpha}} +
\delta _{ap} g_{\kappa \beta} g ^{\lambda \nu} \frac{ \partial
H^{\mu}}{
\partial A_n ^{\alpha}} \right. \nonumber \\
& & \left. \mbox{} - \delta _{an} g ^{\lambda \mu} g_{\alpha}
^{\nu} \frac{ \partial
H_{\kappa}}{ \partial A_p ^{\beta}} + \delta _{an} g ^{\lambda
\nu} g_{\alpha}
^{\mu} \frac{ \partial H_{\kappa}}{ \partial A_p ^{\beta}}
\right\} = 0.
\label{SymX} \end{eqnarray}
Eq.~(\ref{prdelta})
means that Eq.~(\ref{SymX}) should hold whenever the
Yang-Mills equations hold.  To investigate this requirement, we
must substitute
Eqs.\ (\ref{YM0})--(\ref{YM3}) into (\ref{SymX}), and see under
what conditions
the resulting equations vanish identically.  In other words, we
have to
investigate the conditions under which the coefficients of
independent
combinations of derivatives of $A$ vanish.  This is what we
proceed to do.
\subsection*{$\partial A \partial \partial A$ terms}
Eqs.\ (\ref{YM0})--(\ref{YM3}) do not involve terms $\partial
_{\lambda}
\partial _{\mu} A_p ^{\beta}$ with $\lambda$, $\mu$ and $\beta$
all different.
In Eq.~(\ref{SymX}),
therefore, the coefficient of each term $(\partial
^{\kappa} A_n ^{\alpha})(\partial _{\lambda}  \partial _{\mu} A_p
^{\beta})$,
with $\lambda$, $\mu$ and $\beta$ all different, must vanish.
Since $\partial
_{\lambda} \partial _{\mu} = \partial _{\mu} \partial
_{\lambda}$, the
coefficient must be symmetrized in $\lambda$ and $\mu$.  So
$\forall n, p, a,
\kappa, \alpha, \nu$ and $\forall \lambda, \mu, \beta \neq$, we
must have
\begin{eqnarray}
\lefteqn{- 2 \delta _{ap} g_{\kappa} ^{\lambda} g_{\beta} ^{\nu}
\frac{ \partial
H^{\mu}}{ \partial A_n ^{\alpha}} - 2 \delta _{ap} g_{\kappa}
^{\mu} g_{\beta}
^{\nu} \frac{ \partial H^{\lambda}}{ \partial A_n ^{\alpha}} +
\delta _{ap}
g_{\kappa \beta} g ^{\lambda \nu} \frac{ \partial H^{\mu}}{
\partial A_n ^{\alpha}}} \nonumber \\
& & \mbox{} + \delta _{ap} g_{\kappa \beta} g ^{\mu \nu} \frac{
\partial H^{\lambda}}{
\partial A_n ^{\alpha}} + \delta _{an} g ^{\lambda \nu}
g_{\alpha} ^{\mu}
\frac{ \partial H_{\kappa}}{ \partial A_p ^{\beta}} + \delta
_{an} g ^{\mu \nu}
g_{\alpha} ^{\lambda} \frac{ \partial H_{\kappa}}{ \partial A_p
^{\beta}} = 0.
\end{eqnarray}
Letting $n = a \neq p$ yields, $\forall p, \kappa, \alpha, \nu$
and $\forall \lambda, \mu, \beta \neq$
\begin{equation}
(g ^{\lambda \nu} g_{\alpha} ^{\mu} + g ^{\mu \nu} g_{\alpha}
^{\lambda})
\frac{ \partial H_{\kappa}}{ \partial A_p ^{\beta}} = 0.
\label{dadda1} \end{equation}
We can set $\nu = \lambda \neq \mu = \alpha$ and obtain $\forall
p, \kappa,
\beta$:
\begin{equation}
\frac{ \partial H^{\kappa}}{ \partial A_p ^{\beta}} = 0.
\label{D1}\end{equation}
Eq.~(\ref{D1})
is a necessary condition for the $\partial A \partial \partial
A$ terms (with $\lambda$, $\mu$ and $\nu$ all different) to
vanish.  Obviously,
it makes all $\partial A \partial \partial A$ terms in
(\ref{SymX}) vanish.
Therefore, there is no need to substitute the
Yang-Mills equations in those terms.
\subsection*{$\partial \partial A$ terms}
Here again, we begin by looking at terms $\partial _{\lambda}
\partial _{\mu}
A_n ^{\alpha}$ in Eq.~(\ref{SymX})
with $\lambda$, $\mu$ and $\alpha$ all
different.  The (symmetrized) coefficients of these terms must
vanish.  This
means that $\forall n, a, \nu$ and $\forall \lambda, \mu, \alpha
\neq$
\begin{equation}
- 2 \delta _{an} g_{\alpha} ^{\nu} (\partial ^{\lambda} H ^{\mu}
+ \partial
^{\mu} H^{\lambda}) + \delta _{an} (g ^{\lambda \nu} \partial
_{\alpha} H
^{\mu} + g ^{\mu \nu} \partial _{\alpha} H ^{\lambda}) - g^{\nu
\mu} \frac{
\partial \Phi _a ^{\lambda}}{ \partial A_n ^{\alpha}} - g^{\nu
\lambda} \frac{
\partial \Phi _a ^{\mu}}{ \partial A_n ^{\alpha}} = 0.
\label{dda1} \end{equation}
Taking $a \neq n$ and $\nu = \mu$, we find that $\forall a, n
\neq$ and
$\forall \lambda, \alpha \neq$
\begin{equation}
\frac{ \partial \Phi _a ^{\lambda}}{ \partial A_n ^{\alpha}} = 0.
\label{D2} \end{equation}
On the other hand, setting $a = n$ in Eq.~(\ref{dda1}),
we get $\forall
\hat{a}, \nu$ and $\forall \lambda, \mu, \alpha \neq$
\begin{equation}
- 2 g_{\alpha} ^{\nu} (\partial ^{\lambda} H ^{\mu} + \partial
^{\mu}
H^{\lambda}) + g ^{\lambda \nu} \left\{ \partial _{\alpha} H
^{\mu} - \frac{
\partial \Phi _{\hat{a}} ^{\mu}}{ \partial A_{\hat{a}} ^{\alpha}}
\right\} + g
^{\mu \nu} \left\{ \partial _{\alpha} H ^{\lambda} - \frac{
\partial \Phi
_{\hat{a}} ^{\lambda}}{\partial A_{\hat{a}} ^{\alpha}} \right\} =
0.
\label{dda2} \end{equation}
Setting $\nu = \alpha$ yields, $\forall \lambda, \mu \neq$
\begin{equation}
\partial ^{\lambda} H ^{\mu} + \partial ^{\mu} H^{\lambda} = 0,
\label{D3} \end{equation}
while setting $\nu = \lambda$ yields, $\forall \hat{a}$ and
$\forall \mu,
\alpha \neq$
\begin{equation}
\partial _{\alpha} H ^{\mu} - \frac{ \partial \Phi _{\hat{a}}
^{\mu}}{ \partial
A_{\hat{a}} ^{\alpha}} = 0.
\label{D4} \end{equation}
Eqs.\ (\ref{D2}), (\ref{D3}) and (\ref{D4}) are necessary and
sufficient
conditions for $\partial \partial A$ terms with $\lambda$, $\mu$
and $\alpha$
all different to vanish.

Let us now turn to $\partial \partial A$ terms with $\lambda =
\mu = \alpha$.
Since the Yang-Mills equations (\ref{YM0})--(\ref{YM3}) do not
involve such
terms, their coefficients can be set equal to zero.  This yields,
$\forall a,
n, \nu, \hat{\alpha}$
\begin{equation}
- 2 \delta _{an} g_{\hat{\alpha}} ^{\nu} \partial ^{\hat{\alpha}}
H
^{\hat{\alpha}} + \delta _{an} g ^{\hat{\alpha} \nu} \partial
_{\hat{\alpha}} H
^{\hat{\alpha}} + \delta _{an} g _{\hat{\alpha}} ^{\hat{\alpha}}
\partial
^{\nu} H ^{\hat{\alpha}} + g^{\hat{\alpha} \hat{\alpha}} \frac{
\partial \Phi
_a ^{\nu}}{ \partial A_n ^{\hat{\alpha}}}  - g^{\nu \hat{\alpha}}
\frac{
\partial \Phi _a ^{\hat{\alpha}}}{ \partial A_n ^{\hat{\alpha}}}
= 0.
\label{dda3} \end{equation}
Making use of Eqs.\ (\ref{D2}), (\ref{D3}) and (\ref{D4}), and
considering in
turn cases where $a$ is equal to $n$ or not, and where $\nu$ is
equal to
$\hat{\alpha}$ or not, it is not difficult to see that
(\ref{dda3}) holds
identically.

We must now turn to the $\partial \partial A$ terms in
Eq.~(\ref{SymX})
with
two and only two of the indices $\lambda, \mu, \alpha$ equal.
For each value
of $n$, there are 24 such terms.  Since these
second-order partial derivatives are
constrained by the Yang-Mills equations, their coefficients
cannot separately
be set equal to zero.  We have to use the Yang-Mills equations to
eliminate some of the
second-order partial derivatives, and set equal to zero the
coefficients of the remaining independent ones.

So we substitute Eqs.\ (\ref{YM0})--(\ref{YM3}) in (\ref{SymX}),
thereby
eliminating, for every value of $n$, the following derivatives:
$\partial _1
\partial _1 A_n^0$, $\partial _2 \partial _2 A_n^1$, $\partial _1
\partial _1
A_n^2$, and $\partial _1 \partial _1 A_n^3$.  The coefficients of
the remaining
second-order partial derivatives are then extracted, and set
equal to zero.
After minor cancellations, there result the following equations,
which hold
$\forall a, n, \nu$ and for $\hat{\mu}$ and $\hat{\alpha}$ as
indicated.

\vspace{3mm}
\noindent Coefficient of $\partial _{\hat{\mu}} \partial
_{\hat{\mu}} A_n^0$,
with $\hat{\mu} = 2, 3$:
\begin{equation}
\delta _{an} \left\{ - 2 g_0 ^{\nu} \partial ^{\hat{\mu}} H
^{\hat{\mu}} + g
^{\hat{\mu} \nu} \partial _0 H ^{\hat{\mu}} + 2 g_0 ^{\nu}
\partial ^1 H ^1 - g
^{1 \nu} \partial _0 H ^1 \right\} - g ^{\nu \hat{\mu}} \frac{
\partial \Phi _a
^{\hat{\mu}}}{ \partial A_n^0} + g ^{\nu 1} \frac{ \partial \Phi
_a ^1}{
\partial A_n^0} = 0.
\label{dda4}\end{equation}

\noindent Coefficient of $\partial _{\hat{\mu}} \partial
_{\hat{\mu}} A_n^1$,
with $\hat{\mu} = 0, 3$:
\begin{equation}
\delta _{an} \left\{ - 2 g_1 ^{\nu} \partial ^{\hat{\mu}} H
^{\hat{\mu}} + g
^{\hat{\mu} \nu} \partial _1 H ^{\hat{\mu}} + g ^{\hat{\mu}
\hat{\mu}} \left[ -
2 g_1 ^{\nu} \partial ^2 H ^2 + g ^{2 \nu} \partial _1 H ^2
\right] \right\} -
g ^{\nu \hat{\mu}} \frac{ \partial \Phi _a ^{\hat{\mu}}}{
\partial A_n^1} - g
^{\hat{\mu} \hat{\mu}} g ^{\nu 2} \frac{ \partial \Phi _a ^2}{
\partial A_n^1}
= 0.
\label{dda5}\end{equation}

\noindent Coefficient of $\partial _{\hat{\mu}} \partial
_{\hat{\mu}} A_n^2$,
with $\hat{\mu} = 0, 3$:
\begin{equation}
\delta _{an} \left\{ - 2 g_2 ^{\nu} \partial ^{\hat{\mu}} H
^{\hat{\mu}} + g
^{\hat{\mu} \nu} \partial _2 H ^{\hat{\mu}} + g ^{\hat{\mu}
\hat{\mu}} \left[ -
2 g_2 ^{\nu} \partial ^1 H ^1 + g ^{1 \nu} \partial _2 H ^1
\right] \right\} -
g ^{\nu \hat{\mu}} \frac{ \partial \Phi _a ^{\hat{\mu}}}{
\partial A_n^2} - g
^{\hat{\mu} \hat{\mu}} g ^{\nu 1} \frac{ \partial \Phi _a ^1}{
\partial A_n^2}
= 0.
\label{dda6}\end{equation}

\noindent Coefficient of $\partial _{\hat{\mu}} \partial
_{\hat{\mu}} A_n^3$,
with $\hat{\mu} = 0, 2$:
\begin{equation}
\delta _{an} \left\{ - 2 g_3 ^{\nu} \partial ^{\hat{\mu}} H
^{\hat{\mu}} + g
^{\hat{\mu} \nu} \partial _3 H ^{\hat{\mu}} + g ^{\hat{\mu}
\hat{\mu}} \left[ -
2 g_3 ^{\nu} \partial ^1 H ^1 + g ^{1 \nu} \partial _3 H ^1
\right] \right\} -
g ^{\nu \hat{\mu}} \frac{ \partial \Phi _a ^{\hat{\mu}}}{
\partial A_n^3} - g
^{\hat{\mu} \hat{\mu}} g ^{\nu 1} \frac{ \partial \Phi _a ^1}{
\partial A_n^3}
= 0.
\label{dda7}\end{equation}

\noindent Coefficient of $\partial _{\hat{\alpha}} \partial _0
A_n
^{\hat{\alpha}}$, with $\hat{\alpha} = 1, 2, 3$:
\begin{eqnarray}
\lefteqn{\delta _{an} \left\{ - g_{\hat{\alpha}} ^{\nu} (\partial
^{\hat{\alpha}}
H^0 + 2 \partial ^0 H ^{\hat{\alpha}}) + g^{0 \nu} (\partial
_{\hat{\alpha}}
H^{\hat{\alpha}} + 2 \partial ^1 H^1) + g _{\hat{\alpha}}
^{\hat{\alpha}} \partial ^{\nu} H^0 - g^{1 \nu} \partial _0 H^1
\right\}} \nonumber \\
& & \mbox{} - g^{\nu 0} \frac{ \partial \Phi _a ^{\hat{\alpha}}}{
\partial A_n
^{\hat{\alpha}}} - g^{\nu \hat{\alpha}} \frac{ \partial \Phi _a
^0}{ \partial
A_n ^{\hat{\alpha}}} - g^{11} \frac{ \partial \Phi _a ^{\nu}}{
\partial A_n ^0}
+ g^{\nu 1} \frac{ \partial \Phi _a ^1}{ \partial A_n ^0} = 0.
\;\;\;\;\;\; \label{dda8} \end{eqnarray}

\noindent Coefficient of $\partial _{\hat{\alpha}} \partial _1
A_n ^{\hat{\alpha}}$, with $\hat{\alpha} = 0, 2, 3$:
\begin{eqnarray}
\lefteqn{\delta _{an} \left\{ - g_{\hat{\alpha}} ^{\nu} (\partial
^{\hat{\alpha}}
H^1 + 2 \partial ^1 H ^{\hat{\alpha}}) + g^{1 \nu} (\partial
_{\hat{\alpha}}
H^{\hat{\alpha}} + 2 \partial ^2 H^2) + g _{\hat{\alpha}}
^{\hat{\alpha}} \partial ^{\nu} H^1 + g^{2 \nu} \partial _1 H^2
\right\}} \nonumber \\
& & \mbox{} - g^{\nu 1} \frac{ \partial \Phi _a ^{\hat{\alpha}}}{
\partial A_n
^{\hat{\alpha}}} - g^{\nu \hat{\alpha}} \frac{ \partial \Phi _a
^1}{ \partial
A_n ^{\hat{\alpha}}} + g^{22} \frac{ \partial \Phi _a ^{\nu}}{
\partial A_n ^1}
- g^{\nu 2} \frac{ \partial \Phi _a ^2}{ \partial A_n ^1} = 0.
\;\;\;\;\;\; \label{dda9} \end{eqnarray}

\noindent Coefficient of $\partial _{\hat{\alpha}} \partial _2
A_n ^{\hat{\alpha}}$, with $\hat{\alpha} = 0, 1, 3$:
\begin{eqnarray}
\lefteqn{\delta _{an} \left\{ - g_{\hat{\alpha}} ^{\nu} (\partial
^{\hat{\alpha}}
H^2 + 2 \partial ^2 H ^{\hat{\alpha}}) + g^{2 \nu} (\partial
_{\hat{\alpha}}
H^{\hat{\alpha}} + 2 \partial ^1 H^1) + g _{\hat{\alpha}}
^{\hat{\alpha}} \partial ^{\nu} H^2 + g^{1 \nu} \partial _2 H^1
\right\}} \nonumber \\
& & \mbox{} - g^{\nu 2} \frac{ \partial \Phi _a ^{\hat{\alpha}}}{
\partial A_n
^{\hat{\alpha}}} - g^{\nu \hat{\alpha}} \frac{ \partial \Phi _a
^2}{ \partial
A_n ^{\hat{\alpha}}} + g^{11} \frac{ \partial \Phi _a ^{\nu}}{
\partial A_n ^2}
- g^{\nu 1} \frac{ \partial \Phi _a ^1}{ \partial A_n ^2} = 0.
\;\;\;\;\;\; \label{dda10} \end{eqnarray}

\noindent Coefficient of $\partial _{\hat{\alpha}} \partial _3
A_n ^{\hat{\alpha}}$, with $\hat{\alpha} = 0, 1, 2$:
\begin{eqnarray}
\lefteqn{\delta _{an} \left\{ - g_{\hat{\alpha}} ^{\nu} (\partial
^{\hat{\alpha}}
H^3 + 2 \partial ^3 H ^{\hat{\alpha}}) + g^{3 \nu} (\partial
_{\hat{\alpha}}
H^{\hat{\alpha}} + 2 \partial ^1 H^1) + g _{\hat{\alpha}}
^{\hat{\alpha}} \partial ^{\nu} H^3 + g^{1 \nu} \partial _3 H^1
\right\}} \nonumber \\
& & \mbox{} - g^{\nu 3} \frac{ \partial \Phi _a ^{\hat{\alpha}}}{
\partial A_n
^{\hat{\alpha}}} - g^{\nu \hat{\alpha}} \frac{ \partial \Phi _a
^3}{ \partial
A_n ^{\hat{\alpha}}} + g^{11} \frac{ \partial \Phi _a ^{\nu}}{
\partial A_n ^3}
- g^{\nu 1} \frac{ \partial \Phi _a ^1}{ \partial A_n ^3} = 0.
\;\;\;\;\;\; \label{dda11} \end{eqnarray}

We can now investigate the conditions under which
Eqs.\ (\ref{dda4})--(\ref{dda11}) vanish.  Let us first consider
the case where
$a \neq n$.  Making
use of Eq.~(\ref{D2}),
it is easy to see that (\ref{dda4})--(\ref{dda7}) hold
identically, whereas (\ref{dda8})--(\ref{dda11}) hold if and only
if, $\forall
\hat{\alpha}$ and $\forall a, n \neq$
\begin{equation}
\frac{ \partial \Phi _a ^1}{ \partial A_n ^1} = \frac{ \partial
\Phi _a
^{\hat{\alpha}}}{ \partial A_n ^{\hat{\alpha}}} .
\label{dda12} \end{equation}
Let us now turn to the case where $a = n$.  Substituting
(\ref{D4}) into
(\ref{dda4})--(\ref{dda7}), we see that the latter vanish if and
only if,
$\forall \hat{\alpha}$
\begin{equation}
\partial _1 H^1 = \partial _{\hat{\alpha}} H^{\hat{\alpha}} .
\label{D5} \end{equation}
Substituting Eqs.\ (\ref{D3}), (\ref{D4}) and (\ref{D5}) into
(\ref{dda8})--(\ref{dda11}), we see that the latter vanish if and
only if,
$\forall \hat{a}, \hat{\alpha}$
\begin{equation}
\frac{ \partial \Phi _{\hat{a}} ^1}{ \partial A_{\hat{a}} ^1} =
\frac{ \partial
\Phi _{\hat{a}} ^{\hat{\alpha}}}{ \partial A_{\hat{a}}
^{\hat{\alpha}}} .
\label{dda13} \end{equation}
Eqs.\ (\ref{dda12}) and (\ref{dda13}) can be combined in the
following, which
holds $\forall a, n, \hat{\alpha}$:
\begin{equation}
\frac{ \partial \Phi _a ^1}{ \partial A_n ^1} = \frac{ \partial
\Phi _a
^{\hat{\alpha}}}{ \partial A_n ^{\hat{\alpha}}} .
\label{D6} \end{equation}
\subsection*{$\partial A \partial A \partial A$ terms}
Since no such terms appear in the Yang-Mills equations, the
substitution
effected before Eq.~(\ref{dda4})
will not change the coefficients of $\partial
A \partial A \partial A$ terms in (\ref{SymX}).  Owing to
Eq.~(\ref{D1}),
these coefficients identically vanish.
\subsection*{$\partial A \partial A$ terms}
Again, no such terms appear in the Yang-Mills equations.  So we
substitute
Eq.~(\ref{D1})
in the coefficients of $\partial A \partial A$ terms, symmetrize
over the interchange of $(\lambda p \beta)$ with $(\kappa n
\alpha)$ and set
the result to zero.  This yields
\begin{equation}
2 g^{\kappa \lambda} \frac{ \partial ^2 \Phi _a^{\nu}}{ \partial
A_p
^{\beta} \partial A_n ^{\alpha}} - g^{\nu \lambda} \frac{
\partial ^2 \Phi
_a^{\kappa}}{ \partial A_p ^{\beta} \partial A_n ^{\alpha}} -
g^{\nu \kappa}
\frac{ \partial ^2 \Phi_a^{\lambda}}{ \partial A_p ^{\beta}
\partial A_n
^{\alpha}} = 0.
\label{dada1} \end{equation}
Eq.~(\ref{dada1})
holds $\forall \kappa, \lambda, \beta, \alpha, \nu, p, n,
a$.  So we must have
\begin{equation}
\frac{ \partial ^2 \Phi _a^{\nu}}{ \partial A_p ^{\beta} \partial
A_n
^{\alpha}}  = 0.
\label{D7} \end{equation}
That is, all second-order derivatives of $\Phi$ with respect to
$A$ vanish.
\subsection*{$\partial A$ terms}
Here the situation is more complicated.  There are such terms in
the Yang-Mills
equations.  Therefore, the substitution effected before
Eq.~(\ref{dda4})
does
change the coefficients of $\partial A$ terms in (\ref{SymX}).
We recall that
we eliminated the following derivatives: $\partial _1 \partial _1
A_n^0$,
$\partial _2 \partial _2 A_n^1$, $\partial _1 \partial _1 A_n^2$,
and $\partial
_1 \partial _1 A_n^3$.  Taking (\ref{D2}) and (\ref{D4}) into
account, we can
see that for $\hat{\mu} \neq \alpha$ and $\hat{\mu} \neq 0$, the
coefficient of
$ \partial _{\hat{\mu}} \partial _{\hat{\mu}} A_n ^{\alpha}$ in
Eq.~(\ref{SymX}) is given by
\begin{equation}
- 2 \delta _{an} g_{\alpha} ^{\nu} \partial ^{\hat{\mu}}
H^{\hat{\mu}} - \frac{
\partial \Phi _a ^{\nu}}{ \partial A_n ^{\alpha}} = K_{a n
\alpha} ^{\hat{\mu}
\nu} .
\label{da1} \end{equation}
Let $K_{a n \alpha} ^{\hat{\mu} \nu}$ denote the left-hand side
of (\ref{da1})
for any value of the indices.  Then $\partial A$ terms coming
from substitution
of Eqs.\ (\ref{YM0})--(\ref{YM3}) can be written as
\begin{equation}
K_{a c \kappa} ^{1 \nu} C_{cbn} \left\{ 2 A_b ^{\alpha} \partial
_{\alpha} A_n
^{\kappa} - A_b ^{\kappa} \partial _{\alpha} A_n ^{\alpha} - A_{b
\alpha}
\partial ^{\kappa} A_n ^{\alpha} \right\},
\end{equation}
where we have used the fact that, owing to Eq.~(\ref{D5}),
$K_{a c 1} ^{2 \nu}
= K_{a c 1} ^{1 \nu}$. The previous expression can be rearranged
as
\begin{equation}
(\partial _{\lambda} A_n ^{\alpha} ) A_b ^{\mu} K _{ac \kappa}
^{1 \nu} C_{cbn}
( 2 g _{\mu} ^{\lambda} g _{\alpha} ^{\kappa} - g _{\alpha}
^{\lambda} g _{\mu}
^{\kappa} - g ^{\lambda \kappa} g _{\mu \alpha} ) .
\label{da2} \end{equation}

The complete set of $\partial A$ terms can now be obtained by
adding the
explicit ones in Eq.~(\ref{SymX})
to expression (\ref{da2}).  Setting their
coefficients equal to zero and rearranging, we find that $\forall
\lambda, \nu,
\alpha, n, a$
\begin{eqnarray}
\lefteqn{C_{adn} \left\{ 2 g_{\alpha} ^{\nu} \Phi _d ^{\lambda} -
g^{\nu \lambda}
\Phi _{d \alpha} - g _{\alpha} ^{\lambda} \Phi _d ^{\nu} \right\}
+ \delta_{an}
\left\{ - g _{\alpha} ^{\nu} \partial _{\mu} \partial ^{\mu} H
^{\lambda} + \partial ^{\nu} \partial _{\alpha} H ^{\lambda}
\right\}} \nonumber \\
& & \mbox{} + 2 \frac{ \partial}{ \partial A_n ^{\alpha}}
\partial ^{\lambda} \Phi _a
^{\nu} - \frac{ \partial}{ \partial A_n ^{\alpha}} \partial
^{\nu} \Phi _a
^{\lambda} - g ^{\nu \lambda} \frac{ \partial}{ \partial A_n
^{\alpha}}
\partial _{\mu} \Phi _a ^{\mu} \nonumber \\
& & \mbox{} + A_b ^{\mu} \left\{ C_{abn} \left[ - 2 g _{\alpha}
^{\nu} \partial _{\mu}
H^{\lambda} + g_{\mu} ^{\nu} \partial _{\alpha} H ^{\lambda} + g
_{\alpha \mu}
\partial ^{\nu} H ^{\lambda}  - 2 g_{\kappa} ^{\nu} \partial ^1
H^1 ( 2 g_{\mu}
^{\lambda} g_{\alpha} ^{\kappa} - g_{\alpha} ^{\lambda} g_{\mu}
^{\kappa} -
g_{\mu \alpha} g ^{\lambda \kappa}) \right] \right. \nonumber \\
& & \mbox{} + C_{abd} \left[ 2 g_{\mu} ^{\lambda} \frac{ \partial
\Phi _d ^{\nu}}{
\partial A_n ^{\alpha}} - g_{\mu} ^{\nu} \frac{ \partial \Phi _d
^{\lambda}}{
\partial A_n ^{\alpha}} - g ^{\nu \lambda} \frac{ \partial \Phi
_{d \mu}}{
\partial A_n ^{\alpha}} \right] \nonumber \\
& & \left. \mbox{} - C_{cbn} \left[ \frac{ \partial \Phi _a
^{\nu}}{ \partial A_c
^{\kappa}} \right] ( 2 g_{\mu} ^{\lambda} g_{\alpha} ^{\kappa} -
g_{\alpha}
^{\lambda} g_{\mu} ^{\kappa} - g_{\mu \alpha} g ^{\lambda
\kappa}) \right\} =
0.
\label{D8} \end{eqnarray}
\subsection*{No-derivative terms}
There are no-derivative terms in the Yang-Mills equations.
Therefore, the
substitution effected before Eq.~(\ref{dda4})
does change the coefficients of
no-derivative terms in (\ref{SymX}).  Terms coming from the
substitution are
given by
\begin{equation}
K_{an \kappa} ^{1 \nu} C_{nbc} C_{clm} A_l ^{\mu} A_m ^{\kappa}
A_{b \mu} .
\label{no1} \end{equation}
The complete set of no-derivative terms can be obtained by adding
the explicit
ones in Eq.~(\ref{SymX})
to expression (\ref{no1}).  Setting their
coefficients equal to zero and making use of Eq.~(\ref{da1}),
we find that
$\forall \nu, a$
\begin{eqnarray}
\lefteqn{\partial _{\lambda} \partial ^{\lambda} \Phi _a ^{\nu} -
\partial
_{\lambda} \partial ^{\nu} \Phi _a ^{\lambda} + A_b ^{\lambda}
C_{abd} \left\{ 2 \partial _{\lambda} \Phi _d ^{\nu} -
g_{\lambda} ^{\nu} \partial _{\kappa} \Phi _d ^{\kappa} -
\partial ^{\nu} \Phi _{d \lambda} \right\}} \nonumber \\
& & \mbox{} + A_l ^{\mu} A_{b \kappa} \left\{ g_{\mu} ^{\nu}
(C_{abc} C_{cdl} + C_{adc}
C_{cbl}) \Phi _d ^{\kappa} + g_{\mu} ^{\kappa} C_{abc} C_{cld}
\Phi _d ^{\nu}
\right\} \nonumber \\
& & \mbox{} - A_l ^{\mu} A_m ^{\kappa} A_{b \mu} C_{nbc} C_{clm}
\left\{ 2 \delta _{an}
g_{\kappa} ^{\nu} \partial ^1 H ^1 + \frac{ \partial \Phi _a
^{\nu}}{ \partial
A_n ^{\kappa}} \right\} = 0.
\label{D9} \end{eqnarray}

We have now obtained all determining equations associated with
the Yang-Mills
equations.  They are given by Eqs.\ (\ref{D1}), (\ref{D2}),
(\ref{D3}),
(\ref{D4}), (\ref{D5}), (\ref{D6}), (\ref{D7}), (\ref{D8}) and
(\ref{D9}).
They are necessary and sufficient conditions for Eq.~(\ref{SymX})
to hold
whenever the Yang-Mills equations hold.

\section{Solution of Determining Equations}
We now proceed to solve the determining equations.  We first note
that the most
general solution of Eqs.\ (\ref{D1}) and (\ref{D7}) is given by
\begin{equation}
H^{\mu} = H^{\mu} (x^{\lambda})
\label{C1} \end{equation}
and
\begin{equation}
\Phi _a ^{\mu} = \bar{f} ^{\mu} _{a b \kappa} (x^{\lambda}) A_b
^{\kappa} + F_a
^{\mu} (x^{\lambda}) ,
\label{C2} \end{equation}
where $H^{\mu}$, $\bar{f} ^{\mu} _{a b \kappa}$ and $F_a ^{\mu}$
are arbitrary
functions of $x^{\lambda}$.  From Eq.~(\ref{D2}),
we see that $\bar{f} ^{\mu}
_{a b \kappa} = 0$ if $a \neq b$ and $\mu \neq \kappa$.  We can
therefore write
\begin{equation}
\Phi _{\hat{a}} ^{\hat{\mu}} = \sum _{\kappa \neq \hat{\mu}}
\bar{f}
^{\hat{\mu}} _{\hat{a} \hat{a} \kappa} (x^{\lambda}) A_{\hat{a}}
^{\kappa} +
\bar{f} ^{\hat{\mu}} _{\hat{a} b \hat{\mu}} (x^{\lambda}) A_b
^{\hat{\mu}} +
F_{\hat{a}} ^{\hat{\mu}} (x^{\lambda}) ,
\label{C3} \end{equation}

From Eq.~(\ref{D4})
we see that, $\forall \hat{a}$ and $\forall \mu, \kappa
\neq$
\begin{equation}
\partial _{\kappa} H^{\mu} = \bar{f} ^{\mu} _{\hat{a} \hat{a}
\kappa}.
\end{equation}
Thus we can write, $\forall \hat{a}$ and $\forall \mu, \kappa
\neq$
\begin{equation}
\bar{f} ^{\mu} _{\hat{a} \hat{a} \kappa} = {f^{\mu}} _{\kappa},
\end{equation}
where, owing to Eq.~(\ref{D3}),
$f^{\mu \kappa}$ is antisymmetric.

From Eq.~(\ref{D5}),
we see that $\partial _{\hat{\alpha}} H^{\hat{\alpha}}$
is independent of $\hat{\alpha}$, and can therefore be written as
$G$.  From
(\ref{D6}), we see that $\bar{f} ^{\hat{\mu}} _{ab \hat{\mu}}$ is
independent
of $\hat{\mu}$, and can therefore be written as $h_{ab}$.  The
upshot is that
the most general solution of Eqs.\ (\ref{D1}), (\ref{D2}),
(\ref{D3}),
(\ref{D4}), (\ref{D5}), (\ref{D6}) and (\ref{D7}) can be written
as
\begin{equation}
\partial _{\kappa} H^{\mu} = {f^{\mu}}_ {\kappa} + g^{\mu}
_{\kappa} G
\label{C4} \end{equation}
and
\begin{equation}
\Phi _a^{\mu}  = {f^{\mu}} _{\kappa} A_a ^{\kappa} + h_{ab} A_b
^{\mu} + F_a
^{\mu},
\label{C5} \end{equation}
where $f^{\mu \kappa} = - f^{\kappa \mu}$, $h_{ab}$, $F_a ^{\mu}$
and $G$ are
arbitrary functions of $x^{\lambda}$.  Note that
\begin{equation}
\frac{ \partial \Phi _a ^{\mu}}{ \partial A_b ^{\kappa}} = g
_{\kappa} ^{\mu}
h_{ab} + \delta _{ab} {f ^{\mu}} _{\kappa}.
\label{C6} \end{equation}

There remains to satisfy Eqs.\ (\ref{D8}) and (\ref{D9})\@.  We
first substitute Eqs.\ (\ref{C4})--(\ref{C6}) into (\ref{D8}).
After cancellations and rearrangement, we find that $\forall
\lambda, \nu, \alpha, n, a$
\begin{eqnarray}
\lefteqn{A_b ^{\mu} \left\{ 2 g_{\alpha} ^{\nu} g_{\mu}
^{\lambda} - g_{\mu} ^{\nu}
g_{\alpha} ^{\lambda} - g_{\alpha \mu} g^{\lambda \nu} \right\}
\left\{ C_{abn} G + C_{abd} h_{dn} - C_{dbn} h_{ad} + C_{adn}
h_{db} \right\}} \nonumber \\
& & \mbox{} + \delta _{an} \left\{ g_{\alpha} ^{\lambda} \partial
^{\nu} G - g_{\alpha}
^{\nu} \partial _{\mu} (f ^{\lambda \mu} + g^{\mu \lambda} G) + 2
\partial
^{\lambda} {f^{\nu}} _{\alpha} - g^{\lambda \nu} \partial _{\mu}
{f^{\mu}}
_{\alpha} \right\} \nonumber \\
& & \mbox{} + 2 g_{\alpha} ^{\nu} \left\{ C_{adn} F_d ^{\lambda}
+ \partial ^{\lambda}
h_{an} \right\} - g ^{\nu \lambda} \left\{ C_{adn} F_{d \alpha} +
\partial
_{\alpha} h_{an} \right\} - g_{\alpha} ^{\lambda} \left\{ C_{adn}
F_d ^{\nu} +
\partial ^{\nu} h_{an} \right\} = 0.
\end{eqnarray}
Since $f^{\mu \kappa}$, $h_{ab}$, $F_a ^{\mu}$ and $G$ are
functions of
$x^{\lambda}$ only, it is clear that the coefficient of $A_b
^{\mu}$ and the
sum of terms independent of $A$ must separately vanish.  By
considering cases
where $a = n$ and $a \neq n$, we find that necessary and
sufficient conditions
for this are the following: First, $\forall a, b, n$
\begin{equation}
C_{abn} G + C_{abd} h_{dn} - C_{dbn} h_{ad} + C_{adn} h_{db} = 0.
\label{b1} \end{equation}
Furthermore, $\forall \lambda$ and $\forall a, n \neq$
\begin{equation}
C_{adn} F_d ^{\lambda} + \partial ^{\lambda} h_{an} = 0.
\label{b2} \end{equation}
Finally, $\forall \lambda, \nu, \alpha, \hat{a}$
\begin{eqnarray}
\lefteqn{g_{\alpha} ^{\lambda} (\partial ^{\nu} G - \partial
^{\nu} h_ {\hat{a} \hat
{a}}) - g_{\alpha} ^{\nu} (\partial _{\mu} f ^{\lambda \mu} +
\partial ^{\lambda} G - 2 \partial ^{\lambda} h _{\hat{a}
\hat{a}})} \nonumber \\
& & \mbox{} + 2 \partial ^{\lambda} {f^{\nu}} _{\alpha} -
g^{\lambda \nu} (\partial
_{\mu} {f^{\mu}} _{\alpha} + \partial _{\alpha} h _{\hat{a}
\hat{a}}) = 0.
\label{C8} \end{eqnarray}
Note that Eq.~(\ref{b2})
can be written in a form that holds $\forall \lambda,
\hat{a}, n$:
\begin{equation}
C_{\hat{a} dn} F_d ^{\lambda} = - \partial ^{\lambda} h_{\hat{a}
n} + \delta
_{\hat{a} n} \partial ^{\lambda} h_{\hat{a} \hat{a}} .
\label{C7} \end{equation}

In Eq.~(\ref{C8}),
set $\nu = \alpha \neq \lambda$.  There results, $\forall
\lambda, \hat{a}$
\begin{equation}
\partial _{\mu} f ^{\lambda \mu} + \partial ^{\lambda} G - 2
\partial
^{\lambda} h _{\hat{a} \hat{a}} = 0.
\label{b3} \end{equation}
Note that this implies that $\partial ^{\lambda} h_{\hat{a}
\hat{a}}$ is
independent of $\hat{a}$.  Substituting (\ref{b3}) back into
(\ref{C8}) yields
$\forall \lambda, \nu, \alpha, \hat{a}$
\begin{equation}
g_{\alpha} ^{\lambda} (\partial ^{\nu} G - \partial ^{\nu} h_
{\hat{a} \hat
{a}}) + 2 \partial ^{\lambda} {f^{\nu}} _{\alpha} - g^{\lambda
\nu} (\partial
_{\alpha} G - \partial _{\alpha} h _{\hat{a} \hat{a}}) = 0.
\end{equation}
For $\nu = \alpha$, this vanishes identically.  If $\nu \neq
\alpha$, we can
have any of three mutually exclusive cases: (i) $\alpha = \lambda
\neq \nu$;
(ii) $\nu = \lambda \neq \alpha$; (iii) $\nu, \lambda, \alpha
\neq$.  Case (i)
yields $\forall \hat{a}$ and $\forall \nu, \hat{\alpha} \neq$
\begin{equation}
\partial ^{\nu} G - \partial ^{\nu} h_{\hat{a} \hat{a}} + 2
\partial
^{\hat{\alpha}} {f^{\nu}} _{\hat{\alpha}} = 0.
\label{b4} \end{equation}
Case (ii) yields a similar equation.  Finally, case (iii) yields,
$\forall
\lambda, \nu, \alpha \neq$
\begin{equation}
\partial ^{\lambda} {f^{\nu}} _{\alpha} = 0.
\label{b5} \end{equation}
Eqs.\ (\ref{b1}), (\ref{b2}), (\ref{b3}), (\ref{b4}) and
(\ref{b5}) represent
all the conditions on the unknown functions $f^{\mu \kappa}$,
$h_{ab}$, $F_a
^{\mu}$ and $G$ provided by Eq.~(\ref{D8})\@.

We now substitute Eqs.\ (\ref{C4})--(\ref{C6}) in
Eq.~(\ref{D9})\@.
After rearrangement, we find that $\forall \nu, a$
\begin{eqnarray}
\lefteqn{A_l ^{\mu} A_m ^{\kappa} A_b ^{\alpha} \left\{ g_{\alpha
\mu} g_{\kappa} ^{\nu} \left[ C_{abc} C_{cld} h_{dm} + C_{nbc}
C_{clm} ( 2 \delta_{an} G - h_{an}) \right] \right.} \nonumber \\
& & \left. \mbox{} + g_{\mu} ^{\nu} (C_{abc} C_{cml} + C_{amc}
C_{cbl} ) f_{\alpha
\kappa} + g _{\alpha \kappa} g_{\mu} ^{\nu} (C_{abc} C_{cdl} +
C_{adc} C_{cbl})
h_{dm} \right\} \nonumber \\
& & \mbox{} + A_l ^{\mu} A_b ^{\alpha} \left\{ g_{\mu} ^{\nu}
\left[ (C_{abc} C_{cdl} +
C_{adc} C_{cbl}) F_{d \alpha} - C_{alb} \partial _{\kappa}
{f^{\kappa}}
_{\alpha} - C_{ald} \partial _{\alpha} h_{db} \right] \right.
\nonumber \\
& & \left. \mbox{} + g_{\alpha \mu} \left[ C_{abc} C_{cld} F_d
^{\nu} - C_{ald}
\partial ^{\nu} h_{db} \right] + 2 g_{\alpha} ^{\nu} C_{ald}
\partial _{\mu}
h_{db} + C_{alb} (2 \partial _{\mu} {f^{\nu}} _{\alpha} -
\partial ^{\nu}
f_{\mu \alpha}) \right\} \nonumber \\
& & \mbox{} + A_l ^{\mu} \left\{ C_{ald} (2 \partial _{\mu} F_d
^{\nu} - g_{\mu} ^{\nu}
\partial _{\alpha} F_d ^{\alpha} - \partial ^{\nu} F_{d \mu}) +
\delta _{al}
(\partial _{\lambda} \partial ^{\lambda} {f^{\nu}} _{\mu} -
\partial _{\lambda}
\partial ^{\nu} {f^{\lambda}} _{\mu}) + g_{\mu} ^{\nu} \partial
_{\lambda}
\partial ^{\lambda} h_{al} - \partial _{\mu} \partial ^{\nu}
h_{al} \right\}
\nonumber \\
& & \mbox{} + \partial _{\lambda} \partial ^{\lambda} F_a ^{\nu}
- \partial _{\lambda}
\partial ^{\nu} F_a ^{\lambda} = 0.
\label{C9} \end{eqnarray}
The (appropriately symmetrized) coefficients of each power of $A$
must
separately vanish.  Let us consider each of them in turn.

It is not difficult to see that, owing to Eq.~(\ref{b2}),
terms independent of
$A$ identically vanish.  Terms linear in $A$ yield, $\forall \nu,
\mu, a, l$
\begin{equation}
C_{ald} (2 \partial _{\mu} F_d ^{\nu} - g_{\mu} ^{\nu} \partial
_{\alpha} F_d
^{\alpha} - \partial ^{\nu} F_{d \mu}) + \delta _{al} (\partial
_{\lambda}
\partial ^{\lambda} {f^{\nu}} _{\mu} - \partial _{\lambda}
\partial ^{\nu}
{f^{\lambda}} _{\mu}) + g_{\mu} ^{\nu} \partial _{\lambda}
\partial ^{\lambda}
h_{al} - \partial _{\mu} \partial ^{\nu} h_{al} = 0.
\end{equation}
For $a \neq l$, Eq.~(\ref{b2})
implies that this holds identically.  For $a =
l$, we have $\forall \nu, \mu, \hat{a}$
\begin{equation}
\partial _{\lambda} \partial ^{\lambda} {f^{\nu}} _{\mu} -
\partial _{\lambda}
\partial ^{\nu} {f^{\lambda}} _{\mu} + g_{\mu} ^{\nu} \partial
_{\lambda}
\partial ^{\lambda} h_{\hat{a} \hat{a}} - \partial _{\mu}
\partial ^{\nu}
h_{\hat{a} \hat{a}} = 0.
\end{equation}
Setting $\mu = \nu$ and summing immediately yields $\forall
\hat{a}$
\begin{equation}
\partial _{\lambda} \partial ^{\lambda} h_{\hat{a} \hat{a}} = 0,
\label{b6} \end{equation}
whence $\forall \nu, \mu, \hat{a}$
\begin{equation}
\partial _{\lambda} \partial ^{\lambda} {f^{\nu}} _{\mu} -
\partial _{\lambda}
\partial ^{\nu} {f^{\lambda}} _{\mu} - \partial _{\mu} \partial
^{\nu}
h_{\hat{a} \hat{a}} = 0.
\label{b7} \end{equation}

We turn to terms quadratic in $A$ in Eq.~(\ref{C9}).
The coefficient of these
terms, symmetrized under the interchange $(\mu, l)
\leftrightarrow (\alpha,
b)$, must vanish.  A rather lengthy but straightforward
calculation, which we
shall not reproduce here, shows that, owing to (\ref{b1}),
(\ref{b3}),
(\ref{b5}) and the antisymmetry of $f_{\mu \alpha}$, the
resulting equation
reduces to an identity.  Similarly, the coefficient of terms
cubic in $A$,
symmetrized under the sixfold interchange $(\mu, l)
\leftrightarrow (\alpha, b)
\leftrightarrow (\kappa, m)$, vanishes identically.  The upshot
is that Eqs.\
(\ref{b6}) and (\ref{b7}) represent all additional conditions on
the unknown
functions $f^{\mu \kappa}$, $h_{ab}$, $F_a ^{\mu}$ and $G$
provided by
Eq.~(\ref{D9})\@.

We now proceed to solve Eqs.\ (\ref{b1}), (\ref{b2}), (\ref{b3}),
(\ref{b4}),
(\ref{b5}), (\ref{b6}) and (\ref{b7}).  First, let us write
(\ref{b3}),
(\ref{b4}), (\ref{b6}) and (\ref{b7}) in a simpler form.
Consider
Eq.~(\ref{b4})
for the three values of $\hat{\alpha} \neq \nu$.  Summing the
three
resulting equations and remembering that ${f^{\nu}} _{\alpha}$
vanishes if $\nu
= \alpha$, we get $\forall \nu, \hat{a}$
\begin{equation}
3 \partial ^{\nu} G - 3 \partial ^{\nu} h_{\hat{a} \hat{a}} + 2
\partial
^{\alpha} {f^{\nu}} _{\alpha} = 0.
\end{equation}
Comparing with (\ref{b3}), we find that $\forall \nu, \hat{a}$
\begin{equation}
\partial ^{\nu} G + \partial ^{\nu} h_{\hat{a} \hat{a}} = 0.
\label{bb1} \end{equation}
Substituting (\ref{bb1}) into (\ref{b3}), (\ref{b7}) and
(\ref{b4}) and
relabelling yields, $\forall \mu, \nu$
\begin{equation}
\partial _{\lambda} f^{\mu \lambda} + 3 \partial ^{\mu} G = 0,
\label{C10} \end{equation}
\begin{equation}
\partial _{\lambda} \partial ^{\lambda} f^{\nu \mu} - \partial
_{\lambda}
\partial ^{\nu} f^{\lambda \mu} + \partial ^{\nu} \partial ^{\mu}
G = 0,
\label{C11} \end{equation}
and, $\forall \mu, \hat{\lambda} \neq$
\begin{equation}
\partial _{\hat{\lambda}} f^{\mu \hat{\lambda}} + \partial ^{\mu}
G = 0.
\label{bb2} \end{equation}
Substituting (\ref{C10}) in (\ref{C11}) yields $\forall \mu, \nu$
\begin{equation}
\partial _{\lambda} \partial ^{\lambda} f^{\nu \mu} = 2 \partial
^{\nu}
\partial ^{\mu} G.
\end{equation}
Since one side is antisymmetric under the interchange $\nu
\leftrightarrow \mu$
and the other side is symmetric, both sides must vanish.  So we
have, $\forall
\nu, \mu$
\begin{equation}
\partial ^{\nu} \partial ^{\mu} G = 0.
\label{bb3} \end{equation}
Owing to (\ref{b5}), Eqs.\ (\ref{bb1}), (\ref{bb2}) and
(\ref{bb3}) are
equivalent to (\ref{b3}), (\ref{b4}), (\ref{b6}) and (\ref{b7}).

The most general solution of Eq.~(\ref{bb3}) is given by
\begin{equation}
G = d + c_{\mu} x^{\mu},
\label{s1} \end{equation}
where $d$ and $c_{\mu}$ are arbitrary constants.  From
(\ref{b5}), we see that
$f^{\nu \alpha}$ is a function of $x^{\nu}$ and $x^{\alpha}$
only.  From
(\ref{bb2}) and (\ref{s1}) we obtain $\forall \mu, \hat{\lambda}
\neq$
\begin{equation}
\partial _{\hat{\lambda}} f^{\mu \hat{\lambda}} = - \partial
^{\mu} G = -
c^{\mu} .
\end{equation}
This implies that the most general solution for $f^{\mu \lambda}$
is
\begin{equation}
f^{\mu \lambda} = - c^{\mu} x^{\lambda} + c^{\lambda} x^{\mu} +
b^{\mu
\lambda},
\label{s2} \end{equation}
where $b^{\mu \lambda}$ are six arbitrary constants such that
$b^{\mu \lambda}
= - b^{\lambda \mu}$.

We can now solve for the functions $H^{\mu} (x^{\lambda})$.  With
(\ref{s1})
and (\ref{s2}), Eq.~(\ref{C4}) can easily be integrated to give
\begin{equation}
H^{\mu} = - \frac{1}{2} c^{\mu} x^{\lambda} x_{\lambda} +
c^{\lambda} x^{\mu}
x_{\lambda} + b^{\mu \lambda} x_{\lambda} + d x^{\mu} + a^{\mu},
\label{s3} \end{equation}
where $a^{\mu}$ are four arbitrary constants.

There remains to solve Eqs.\ (\ref{b1}), (\ref{b2}) and
(\ref{bb1})\@.  In
Appendix B, we shall show by group theoretical arguments that the
most general
solution of Eq.~(\ref{b1}) is given by
\begin{equation}
h_{ab} = - G \delta_{ab} + C_{abd} \chi _d ,
\label{s4} \end{equation}
where the $\chi _d$ are arbitrary functions of $x^{\lambda}$.
Substituting
Eq.~(\ref{s4})
in (\ref{bb1}), we see that the latter holds identically.
Substituting (\ref{s4}) in (\ref{b2}), we find that $\forall
\lambda$ and
$\forall a, n \neq$
\begin{equation}
C_{adn} F_d ^{\lambda} = - \partial ^{\lambda} C_{and} \chi _d ,
\end{equation}
whence, owing to Eq.~(\ref{Kro})
and the antisymmetry of the structure
constants
\begin{equation}
F_d ^{\lambda} = \partial ^{\lambda} \chi _d .
\label{s5} \end{equation}
Putting together Eqs.\ (\ref{C5}), (\ref{s1}), (\ref{s2}),
(\ref{s4}) and
(\ref{s5}), we find that
\begin{equation}
\Phi _a^{\mu} = (- c^{\mu} x_{\lambda} + c_{\lambda} x^{\mu} +
{b^{\mu}}
_{\lambda}) A_a ^{\lambda} - (d + c_{\lambda} x^{\lambda}) A_a
^{\mu} + C_{abd}
\chi _d A_b ^{\mu} + \partial ^{\mu} \chi _a .
\label{s6} \end{equation}

Eqs.\ (\ref{s3}) and (\ref{s6}) are the most general solution of
the
determining equations.  Therefore, the corresponding vector field
(\ref{v})
generates Lie symmetries of the Yang-Mills equations.  One can
see that the
constants $a^{\mu}$ correspond to space-time translations; that
the $b^{\lambda
\mu}$ correspond to Lorentz transformations; that the $c^{\mu}$
correspond to
uniform accelerations; that $d$ corresponds to dilatations; and
that the
functions $\chi _a (x^{\lambda})$ correspond to local gauge
transformations~\cite{Wess,Salam}.
We have thus recovered the well-known Lie symmetries of the
Yang-Mills equations.  But we have done much more.
Isofar as the Yang-Mills equations are locally
solvable, we have shown that there are no others.

\section{Gauge Conditions}
In Eqs.\ (\ref{s3}) and (\ref{s6}), we have obtained the
coefficients of
symmetry generators of the Yang-Mills equations.  In practice,
the equations
will be used together with a gauge condition.  So it is of
interest to
investigate the symmetries of the Yang-Mills equations in a
particular gauge.
To be specific, we shall pick the Lorentz gauge.

The Lorentz gauge condition consists in setting $\forall a$
\begin{equation}
\partial _{\mu} A_a ^{\mu} = 0 .
\label{Lorentz} \end{equation}
Our task consists in finding the Lie symmetries of Eqs.\
(\ref{YMX}) and
(\ref{Lorentz})\@.

It is not difficult to check that Eqs.\ (\ref{YMX}) and
(\ref{Lorentz})
together have maximal rank.  But they are not locally solvable.
Differentiating (\ref{Lorentz}) with respect to $x^{\lambda}$, we
find that
\begin{equation}
\partial _{\lambda} \partial _{\mu} A_a ^{\mu} = 0 ,
\label{cons} \end{equation}
which are additional constraints on partial derivatives.

It is shown in Ref.~\cite{Olver}
that a necessary and sufficient condition
for $v$ to
generate a symmetry of a system of {\it n}-th order equations is
that the {\it
n}-th prolongation of $v$, acting on the system, vanishes at all
points where
the system is locally solvable.  In our case, such points are
determined by
Eqs.\ (\ref{YMX}), (\ref{Lorentz}), (\ref{cons}), and any other
equation
expressing constraints on the $A_a ^{\mu}$ and their first and
second-order derivatives.  For similar reasons as given in
Section~2, however,
it is likely that there are no additional constraints.  We shall
thus
investigate the conditions under which the second prolongation of
$v$, acting
on Eqs.\ (\ref{YMX}) and (\ref{Lorentz}), vanish whenever Eqs.\
(\ref{YMX}),
(\ref{Lorentz}) and (\ref{cons}) hold.

Let us apply the second prolongation operator (\ref{pr2}) to
Eqs.\ (\ref{YMX})
and (\ref{Lorentz}), and set the result to zero.  Applying
(\ref{pr2}) to
Eq.~(\ref{YMX}), we clearly recover Eq.~(\ref{SymX})\@.
Applying (\ref{pr2}) to
(\ref{Lorentz}), we find that
\begin{equation}
\Phi _{a \mu} ^{\mu} = 0 ,
\end{equation}
or, using (\ref{Phi3X})
\begin{equation}
\partial _{\mu} \Phi _a^{\mu}  - (\partial _{\mu} H^{\nu})
\partial _{\nu} A_a
^{\mu} + (\partial _{\mu} A_n ^{\alpha}) \frac{\partial}{\partial
A_n
^{\alpha}} \Phi _a^{\mu} - (\partial _{\mu} A_n ^{\alpha})
(\partial _{\nu} A_a
^{\mu}) \frac{\partial}{\partial A_n ^{\alpha}} H^{\nu} = 0 .
\label{SymL} \end{equation}

We now have to substitute Eqs.\ (\ref{YMX}), (\ref{Lorentz}), and
(\ref{cons})
into (\ref{SymX}) and (\ref{SymL}), and equate to zero the
coefficients of the
remaining (independent) combinations of derivatives of $A$.  Note
that this
complicated and correct procedure is not the same as the simpler
one that
consists in substituting Eqs.\ (\ref{s3}) and (\ref{s6}) into
(\ref{SymL}),
although in specific instances the two procedures may yield the
same results.

Let us then consider in turn the various combinations of
derivatives of $A$.
The $\partial A \partial \partial A$ terms can be treated
basically as in
Section~2\@.  We recall that only terms $\partial _{\lambda}
\partial _{\mu}
A_p ^{\beta}$, with $\lambda$, $\mu$ and $\beta$ all different,
had to be
considered.  Thus, substitution of (\ref{cons}) will not have any
effect.
Moreover, it is not difficult to see that Eq.~(\ref{D1})
still obtains if we
restrict our attention to terms $\partial ^{\kappa} A_n
^{\alpha}$ with $\kappa
\neq \alpha$.  But the Lorentz gauge condition does not involve
such terms.
Eqs.\ (\ref{D1}), therefore, are still necessary and sufficient
conditions for
the $\partial A \partial \partial A$ terms to vanish.

Discussion of $\partial \partial A$ terms is not much changed
either.
Eq.~(\ref{cons})
allows to write terms like $\partial _{\hat{\alpha}} \partial
_{\hat{\alpha}} A_a ^{\hat{\alpha}}$ in terms of other
second-order derivatives of $A$.  But the coefficient of
$\partial
_{\hat{\alpha}} \partial _{\hat{\alpha}} A_a ^{\hat{\alpha}}$ is
given by the
left-hand side of Eq.~(\ref{dda3}),
which was shown to vanish identically.  So
again, the substitution of the Lorentz gauge condition and its
derivatives will
not introduce anything new.

It is easy to see that $\partial A \partial A \partial A$ terms
still vanish
identically.  Turning to $\partial A \partial A$ terms, we can
see that
Eq.~(\ref{D7})
can be obtained even if we restrict our attention to terms with
$\lambda \neq \beta$ and $\kappa \neq \alpha$.  The $\partial A$
terms yield
Eqs.\ (\ref{b1}), (\ref{b2}), (\ref{b3}), (\ref{b4}) and
(\ref{b5}) even if we
restrict ourselves to $\lambda \neq \alpha$.  Finally, terms with
no
derivatives of $A$ do not change, since the Lorentz gauge
condition involves
derivatives only.

The upshot of the foregoing analysis is that the conditions that
make
(\ref{SymX}) vanish subject to (\ref{YMX}), (\ref{Lorentz}), and
(\ref{cons})
are the same as the ones that make (\ref{SymX}) vanish subject to
(\ref{YMX})
only.  In the end, these conditions are precisely embodied in
Eqs.\ (\ref{s3})
and (\ref{s6}).  We stress that this is not obvious, and could be
otherwise for
other choices of gauge.

There remains to make use of Eq.~(\ref{SymL})
to put further constraints on
the functions $H^{\mu}$ and $\Phi _a^{\mu}$.  Substituting
(\ref{s3}) and
(\ref{s6}) into (\ref{SymL}) and rearranging, we find that
\begin{equation}
2 c_{\kappa} A_a ^{\kappa} + C_{abd} (\partial _{\mu} \chi _d)
A_b ^{\mu} +
\partial _{\mu} \partial ^{\mu} \chi _a = 0 .
\end{equation}
This must hold identically.  Since $\chi _a$ is a function of
$x^{\lambda}$
only, we get
\begin{equation}
\partial _{\mu} \partial ^{\mu} \chi _a = 0 ,
\end{equation}
\begin{equation}
2 c_{\mu} \delta _{ab} + C_{abd} \partial _{\mu} \chi _d = 0.
\end{equation}
Necessary and sufficient conditions for these two equations to
hold are that
$(\forall \mu)$,
$c_{\mu} = 0$ and that $(\forall \mu, d)$, $\partial
_{\mu} \chi
_d = 0$.  The conformal symmetry thus collapses to the
Poincar\'{e} group with
dilatations, and local gauge transformations reduce to global
ones.

\section*{Acknowledgements}
I am thankful to A. M. Grundland for numerous discussions on the
symmetries of
differential equations.  This work was supported by the Natural
Sciences and
Engineering Research Council of Canada.
\appendix
\section{}
To see whether there are additional constraints on the $A_a
^{\nu}$ and their
derivatives, let us apply the operator $\partial _{\nu}$ on
Eq.~(\ref{YMX})\@.
We get
\begin{eqnarray}
\lefteqn{C_{abc} \left\{ 2 A_b ^{\mu} \partial _{\nu} \partial
_{\mu} A_c ^{\nu} + 2
(\partial _{\nu} A_b ^{\mu}) \partial _{\mu} A_c ^{\nu} +
(\partial _{\nu} \partial _{\mu} A_b ^{\mu}) A_c ^{\nu} +
(\partial _{\mu} A_b ^{\mu}) \partial
_{\nu} A_c ^{\nu} - (\partial _{\nu} A_{b {\mu}}) \partial ^{\nu}
A_c ^{\mu} \right.} \nonumber \\
& & \left. \mbox{} - A_{b \mu} \partial _{\nu} \partial ^{\nu}
A_c ^{\mu} + C_{cdl}
\left[ A_{b \mu} A_l ^{\nu} \partial _{\nu} A_d ^{\mu} +  A_{b
\mu} A_d ^{\mu}
\partial _{\nu} A_l ^{\nu} + A_d ^{\mu} A_l ^{\nu} \partial
_{\nu} A_{b \mu}
\right] \right\} = 0.
\end{eqnarray}
The second, fourth, and fifth terms in curly brackets vanish due
to
antisymmetry of $C_{abc}$.  The third term similarly cancels half
the first
term.  Thus we obtain
\begin{equation}
C_{abc} \left\{ A_b ^{\mu} \partial _{\nu} \partial _{\mu} A_c
^{\nu} - A_{b
\mu} \partial _{\nu} \partial ^{\nu} A_c ^{\mu} + C_{cdl} \left[
A_{b \mu} A_l
^{\nu} \partial _{\nu} A_d ^{\mu} +  A_{b \mu} A_d ^{\mu}
\partial _{\nu} A_l
^{\nu} + A_d ^{\mu} A_l ^{\nu} \partial _{\nu} A_{b \mu} \right]
\right\} = 0.
\end{equation}
Substituting Eq.~(\ref{YMX})
and again making use of the antisymmetry of the
structure constants, we get
\begin{equation}
C_{abc} C_{cdl} \left\{ A_{b {\mu}} A_d ^{\nu} \partial _{\nu}
A_l ^{\mu} -
A_{b {\mu}} A_{d {\nu}} \partial ^{\mu} A_l ^{\nu} + A_d ^{\mu}
A_l ^{\nu}
\partial _{\nu} A_{b \mu} + C_{lmn} A_{b \mu} A_m ^{\nu} A_n
^{\mu} A_{d \nu}
\right\} = 0.
\end{equation}
Relabeling indices in the second and last terms and regrouping
yields
\begin{eqnarray}
\lefteqn{\left\{ C_{abc} C_{cdl} + C_{adc} C_{clb} \right\} A_{b
{\mu}} A_d ^{\nu}
\partial _{\nu} A_l ^{\mu} + C_{abc} C_{cdl} A_d ^{\mu} A_l
^{\nu} \partial _{\nu} A_{b \mu}} \nonumber \\
& & \mbox{} + \frac{1}{2} \left\{ C_{abc} C_{cdl} + C_{adc}
C_{clb} \right\} C_{lmn}
A_{b \mu} A_m ^{\nu} A_n ^{\mu} A_{d \nu} = 0.
\end{eqnarray}
Making use of the Jacobi identities for the structure constants,
we get
\begin{equation}
- C_{alc} C_{cbd} A_{b {\mu}} A_d ^{\nu} \partial _{\nu} A_l
^{\mu} + C_{abc}
C_{cdl} A_d ^{\mu} A_l ^{\nu} \partial _{\nu} A_{b \mu} -
\frac{1}{2} C_{alc}
C_{cbd} C_{lmn} A_{b \mu} A_m ^{\nu} A_n ^{\mu} A_{d \nu} = 0.
\end{equation}
The first two terms cancel and, by antisymmetry of the structure
constants, the
third term vanishes.
\section{}
We want to solve Eq.~(\ref{b1}), namely
\begin{equation}
C_{abn} G + C_{abd} h_{dn} - C_{dbn} h_{ad} + C_{adn} h_{db} = 0.
\label{a1} \end{equation}
The $C_{abn}$ are structure constants of a compact semisimple Lie
group, the
function $G$ is given by Eq.~(\ref{s1})
and the $h_{ad}$ are unknown functions
of $x^{\lambda}$.

We fix the value of $x^{\lambda}$, so that $G$ and $h_{ad}$ are
fixed too.  In
(\ref{a1}), we interchange $a$ with $n$, and add the result to
(\ref{a1}).  We
obtain
\begin{equation}
C_{abd} (h_{dn} + h_{nd}) + C_{nbd} (h_{da} + h_{ad}) = 0
\end{equation}
or, in matrix notation
\begin{equation}
\left[ C_b , h + h^T \right] = 0 ,
\label{a2} \end{equation}
where $C_b$ has elements $C_{bad}$, $h$ has elements $h_{ad}$ and
$h^T$ is the
transpose of $h$.  Now the structure constants are matrices of an
irreducible
representation of the Lie algebra.  By Schur's lemma,
Eq.~(\ref{a2})
implies
that $h + h^T$ is a multiple of the identity, that is,
\begin{equation}
h_{ad} + h_{da} = 2 \lambda \delta _{ad} .
\label{a3} \end{equation}

Let us denote by $M_{ad}$ the antisymmetric part of $h_{ad}$.
Owing to
(\ref{a3}), Eq.~(\ref{a1}) becomes
\begin{equation}
C_{abn} G + C_{abd} (\lambda \delta_{dn} + M_{dn}) - C_{dbn} (
\lambda
\delta_{ad} + M_{ad}) + C_{adn} (\lambda \delta _{db} + M_{db}) =
0,
\end{equation}
which reduces to
\begin{equation}
C_{abn} (G + \lambda) + C_{abd} M_{dn} + C_{bnd} M_{da} + C_{nad}
M_{db} = 0.
\label{a4} \end{equation}

We multiply this equation by $C_{abl}$, sum over $a$ and $b$ and
make use of
Eq.~(\ref{Kro}) to obtain
\begin{equation}
\delta _{ln} (G + \lambda) + M_{ln} + 2 C_{abl} C_{bnd} M_{da} =
0 .
\label{a45} \end{equation}
The first term is symmetric under the interchange $l
\leftrightarrow n$,
whereas the last two terms are antisymmetric.  This means that
\begin{equation}
G + \lambda = 0,
\label{a5} \end{equation}
whence
\begin{equation}
h_{ad} = - G \delta _{ad} + M_{ad} .
\label{a6} \end{equation}
Eq.~(\ref{a4}) becomes
\begin{equation}
C_{abd} M_{dn} + C_{bnd} M_{da} + C_{nad} M_{db} = 0.
\label{a7} \end{equation}

It is obvious that, for any set of $\chi _l$, the following is a
solution of
Eq.~(\ref{a7}):
\begin{equation}
M_{dn} = C_{dnl} \chi _l .
\label{a8} \end{equation}
We shall now show that there are no other solutions.

Owing to (\ref{a5}), Eq.~(\ref{a45})
can be written in matrix form as
\begin{equation}
M + 2 \sum _b C_b M C_b = 0 .
\label{a9} \end{equation}
Let $L$ denote the Lie algebra whose structure constants are the
$C_{bad}$.
Then $L$ is semisimple.  In a suitable basis, each matrix $C_b$
is block
diagonal, with nonzero entries in one block only.  Each block
corresponds to a
simple subalgebra of $L$.  From Eq.~(\ref{a9}),
it follows that $M$ is also
block diagonal.  Eq.~(\ref{a9}),
therefore, holds separately for each block.
Thus it is enough to consider the case where $L$ is simple.

From (\ref{a7}), we have
\begin{equation}
\left[ C_a, M \right] = \sum _d M_{ad} C_d .
\label{a10} \end{equation}
Suppose there is a matrix $M$ that satisfies (\ref{a7}) and is
not a linear
combination of the $C_d$.  From (\ref{a10}), we see that the
$C_d$ and $M$
together form a Lie algebra that includes $L$.  Let $N$ denote
the dimension of
$L$, and let M($L;R$) denote the real irreducible representation
of $L$ made up
of the structure constants.  Let M($L;C$) denote the
corresponding
representation of the complex form of $L$.  It is known that
M($L;C$) is
maximal in the orthogonal algebra SO($N;C$)~\cite{Dynkin}.
From this it
follows that M($L;R$) is maximal in SO($N;R$)\@.  For, if there
existed a real
Lie algebra $L'$ such that
\begin{equation}
\mbox{M} (L;R) \subset L' \subset \mbox{SO} (N;R),
\end{equation}
corresponding inclusions would also hold for the complex forms.
But
\begin{equation}
\mbox{dim} \left\{ \mbox{SO} (N;R) \right\}  - \mbox{dim} \left\{
L \right\} =
\frac{N(N-3)}{2}.
\end{equation}
Since this is never equal to 1, the $C_d$ and $M$ cannot together
form a Lie
algebra.  The upshot is that Eq.~(\ref{a8})
is the most general solution of
(\ref{a7}).  Eq.~(\ref{a6}) thus becomes
\begin{equation}
h_{ad} = - G \delta _{ad} + C_{abd} \chi _d.
\label{a11} \end{equation}
Since this holds at any point $x^{\lambda}$, Eq.~(\ref{s4})
follows.


\begin{thebibliography}{99}
\bibitem{Olver}P. J. Olver, {\it Applications of Lie Groups to
Differential
Equations} (Springer-Verlag, New York, 1986).
\bibitem{Torrea}C. G. Torre and I. M. Anderson, Phys.\ Rev.\
Lett.\ {\bf 70}, 3525 (1993).
\bibitem{Rosen}V. Rosenhaus and K. Kiiranen, Proc.\
Acad.\ Sc.\ Estonian SSR:
Phys.\ Math.\ {\bf 31}, 304 (1982).
\bibitem{Ker}P. H. M. Kersten, {\it Infinitesimal Symmetries: A
Computational
Approach}, Ph. D. thesis (Twente University of Technology,
Enschede, The
Netherlands, 1985); CWI Tract 34 (Center for Mathematics and
Computer Science,
Amsterdam, 1987).
\bibitem{Torre}C. G. Torre, J. Math.\ Phys.\ {\bf 36}, 2113
(1995).
\bibitem{Abers}E. S. Abers and B. W. Lee, Phys.\
Reports {\bf9}, 1 (1973).
\bibitem{Huang}K. Huang, {\it Quarks, Leptons and Gauge Fields}
(World
Scientific, Singapore, 1992).
\bibitem{Wess}J. Wess, Il Nuovo Cimento {\bf18}, 1086 (1960).
\bibitem{Salam}G. Mack and A. Salam, Ann.\
Phys.\ (NY) {\bf53}, 174 (1969).
\bibitem{Dynkin}E. B. Dynkin, Am.\ Math.\ Soc.\ Translations
{\bf6}, Series 2, 245 (1957).
\end{thebibliography}
\end{document}